\documentclass[11pt]{article}

\usepackage{amsmath}

\begin{document}

\title{Cryptanalysis of Song's advanced smart card based password authentication protocol}

\author{Juan E. Tapiador$^{1,\ast}$, Julio C. Hernandez-Castro$^2$\\Pedro Peris-Lopez$^3$, John A. Clark$^1$,\\
\footnotesize{\emph{$^1$ Department of Computer Science, University of York, UK}}\\
\footnotesize{$^{\ast}$ Corresponding author. E-mail: \texttt{jet@cs.york.ac.uk}}\\
\footnotesize{\emph{$^2$ School of Computing, University of Portsmouth, UK}}\\
\footnotesize{\emph{$^3$ Security Lab, Faculty of EEMCS, Delft University of Technology, The Netherlands}}
}

\maketitle

\begin{abstract}
Song \cite{Song10} proposed very recently a password-based authentication and key establishment protocol using smart cards which attempts to solve some weaknesses found in a previous scheme suggested by Xu, Zhu, and Feng \cite{XZF09}. In this paper, we present attacks on the improved protocol, showing that it fails to achieve the claimed security goals.
\end{abstract}

\section{Introduction}
Remote user authentication is a central problem in network security. In a seminal paper, Lamport \cite{Lamport81} proposed in 1981 a password-based scheme using hash chains. This scheme was later refined and used in a number of applications, notably Haller's famous S/KEY one-time password system \cite{Haller94}. Similar protocols based on smart cards gained some popularity shortly after that. In such schemes, the user is provided with a card and a password as identification tokens. When the user wishes to connect to the server, she provides the card with her password, which is used to construct a login message that is sent to the server to be validated. More sophisticate schemes force the server to be authenticated too, and also provide both parties with a shared secret (a session key) after the completion of the protocol.

The common adversary model to analyze the security of authentication protocols based on smart cards assumes an attacker with full control over the communication channel between the user and the server.
Consequently, all the messages exchanged can be intercepted, deleted, modified, or fabricated by the attacker. Additionally, protocols must assume that the attacker can temporarily get access to the user's smart card and the information stored in it, either directly (e.g. stealing the card or deceiving the user so she inserts the card in a malicious reader) or indirectly by observing emanations or other side channels \cite{KFD,MDS}.

Very recently, Song \cite{Song10} showed various attacks against one of such protocols suggested by Xu, Zhu, and Feng \cite{XZF09}. The paper also presents an improved version, loosely based on the original scheme, which attempts to amend the identified vulnerabilities. In particular, Song claims that \cite{Song10}: \emph{``The interactive authentication messages must not reduce the entropy of the password''}, and also: \emph{``The adversary must not be able to attack and gain access to the system by extracting the data stored on the smart card''}. In this paper, we present practical attacks showing that the protocol suggested by Song fails to achieve these goals.

\section{Review of Song's scheme}
We first give a brief description of Song's scheme as presented in \cite{Song10}. The notation used in the protocol is summarized in Fig. \ref{Fig:Notation}.

\begin{figure}[t]
\begin{center}
\begin{tabular}{|ll|}
\hline
$S$ & Server\\
$A$ & User\\
$ID_A$ & User $A$'s identity\\
$PW_A$ & User $A$'s password\\
$R_A$ & One-time random number generated by $A$\\
$T_A, T_S$ & User $A$ and server's timestamps, respectively\\
$\Delta T$ & Time threshold predefined by the protocol\\
$p, q$ & Large prime numbers such that $p=2q + 1$\\
$x \in Z_q^{\ast}$ & Server's secret key\\
$\oplus$ & Bitwise XOR operation\\
$\parallel$ & Concatenation operation\\
$h(\cdot)$ & A secure one-way hash function\\
$E_K(M), D_K(M)$ & Encryption/Decryption of message $M$ with key $K$\\
\hline
\end{tabular}
\end{center}
\caption{Notation used in Song's protocol.}\label{Fig:Notation}
\end{figure}

Initially, the server selects two large prime numbers $p$ and $q$ such that $p=2q + 1$, and a secret key $x \in Z_q^{\ast}$. Both $p$ and $x$ are kept secret. The protocol consists of four main phases (see Fig. \ref{Fig:Protocol}).

\begin{figure}[t]
\begin{footnotesize}
\begin{center}
\begin{tabular}{|lcl|}
\hline
User $A$ & & \multicolumn{1}{r|}{Server $S$}\\
\hline
\multicolumn{3}{|c|}{Registration phase}\\
Select $ID_A, PW_A$ & $\xrightarrow{ID_A,PW_A}$ & $B_A=h(ID^x~\textrm{mod}~p) \oplus h(PW_A)$\\
& $\xleftarrow{\textrm{Smart~card}}$  & Store ${ID_A, B_A}$ in the card\\
\hline
\multicolumn{3}{|c|}{Login and authentication}\\
Input $ID_A, PW_A$ & & \\
Select $R_A$& & \\
$K_A = B_A \oplus h(PW_A)$ & & \\
$W_A = E_{K_A}(R_A \oplus T_A)$  & & \\
$C_A = h(T_A \parallel R_A \parallel W_A \parallel ID_A)$  & & \\
& $\xrightarrow{ID_A, C_A, W_A, T_A}$ & Verify $ID_A, T_A$\\
& & $K_A = h(ID^x~\textrm{mod}~p)$ \\
& & $R_A' = D_{K_A}(W_A) \oplus T_A$\\
& & $C_A' = h(T_A \parallel R_A' \parallel W_A \parallel ID_A)$\\
& & Verify: $C_A \stackrel{?}{=} C_A'$\\ 
& & $C_S = h(ID_A \parallel R_A' \parallel T_S)$\\
& $\xleftarrow{ID_A, C_S, T_S}$ & \\
Verify $ID_A, T_S$ & & \\
$C_S' = h(ID_A \parallel R_A \parallel T_S)$ & &\\
Verify $C_S \stackrel{?}{=} C_S'$ & &\\
\hline
\multicolumn{3}{|c|}{Compute session key}\\
$sk = h(ID_A \parallel T_S \parallel T_A \parallel R_A)$ & &
$sk = h(ID_A \parallel T_S \parallel T_A \parallel R_A')$\\
\hline
\end{tabular}
\end{center}
\end{footnotesize}
\caption{Song's protocol.}\label{Fig:Protocol}
\end{figure}

\subsection{Registration phase}
The user $A$ sends to $S$ her identity $ID_A$ and password $PW_A$ through a secure channel. The server then computes $B_A=h(ID^x~\textrm{mod}~p) \oplus h(PW_A)$, stores both $ID_A$ and $B_A$ in a smart card and sends it to $A$.

\subsection{Login phase}
User $A$ attachs her smart card to a reader and enters her identity and password. The card chooses a random number $R_A$, obtains the current timestamp $T_A$, and computes:
$$K_A = B_A \oplus h(PW_A)$$
$$W_A = E_{K_A}(R_A \oplus T_A)$$
$$C_A = h(T_A \parallel R_A \parallel W_A \parallel ID_A)$$
It then sends the login message $\{ID_A, C_A, W_A, T_A\}$ to the server.

\subsection{Authentication phase}

\subsubsection{User Authentication}
Upon receiving the login request at time $T^{\ast}$, $S$ first checks $A$'s identity and then validates the timestamp by checking that $(T^{\ast} - T_A) \leq \Delta T$. The server computes a local version of the session key as $K_A=h(ID^x~\textrm{mod}~p)$ and then recovers the nonce by doing $R_A = D_{K_A}(W_A) \oplus T_A)$. It then computes a local version of $C_A$ and checks whether it coincides with the received value. If the verification goes through successfully, the user is authenticated and $S$ sends her the message $\{ID_A, C_S, T_S\}$, where $T_S$ is the server's timestamp and $C_S=h(ID_A \parallel R_A \parallel T_S)$.

\subsubsection{Server authentication}
Upon receviving the server's last message, $A$ validates the identity and the timestamp, and verifies that the received $C_S$ coincides with a local version computed by her using the original nonce. If that is the case, then $S$ is authenticated.

\subsubsection{Session key establishment}
Once both $A$ and $S$ are mutually authenticated, they compute a shared secret session key $sk = h(ID_A \parallel T_S \parallel T_A \parallel R_A)$, which is used to encrypt future communications.

\subsection{Password change}
Whenever the user wants to change her password, she first goes through the authentication protocol. Upon receving the successful authentication confirmation from the server, $A$ introduces her new password $PW_A^{new}$ and the smart card updates the value of $B_A$ by doing\footnote{We note that the actual formulation of the update process described in \cite{Song10} is $B_A^{new} = B_A \oplus PW_A \oplus PW_A^{new}$, which is clearly erroneous.} $B_A^{new} = B_A \oplus h(PW_A) \oplus h(PW_A^{new})$.

\section{Cryptanalysis}

\subsection{Off-line password guessing attack}

In \cite{Song10} it is claimed that \emph{``the adversary must not be able to attack and gain access to the system by extracting the data stored on the smart card.''} However, an adversary who obtains the value $B_A = h(ID^x~\textrm{mod}~p) \oplus h(PW_A)$ can easily mount an off-line password guessing attack by simply observing one correct authentication session and getting access to the values $W_A$ and $C_A$.

The attack works as follows. For each candidate password $PW_A^{\ast}$, the attacker computes the tentative encryption key $K_A^{\ast} = B_A \oplus h(PW_A^{\ast})$. Such a key is then used to recover the candidate nonce value $R_A^{\ast}$ by first decrypting $W_A$ with $K_A^{\ast}$ and then XORing the result with $T_A$ (both of which are public); that is, $R_A^{\ast} = D_{K_A^{\ast}}(W_A) \oplus T_A$. Note that, if the attempted password $PW_A^{\ast}$ is correct (i.e., $PW_A^{\ast} = PW_A$), then so it is the obtained encryption key $K_A^{\ast}$ and, consequently, the nonce $R_A^{\ast}$. Now, $C_A$ can be used to check if that is the case: The attacker computes $C_A^{\ast}=h(T_A \parallel R_A^{\ast} \parallel W_A \parallel ID_A)$ and, if it coincides with $C_A$, she can conclude that $R_A^{\ast}$ is correct and so the candidate password tried. (In this reasoning we assume that $h$ has no collisions. Nevertheless, even if $h$ is not ideal, additional eavesdropped sessions can be used to rule out false positives and identify the correct password).

In short, contrarily to what is claimed in \cite{Song10}, the messages exchanged during the protocol do indeed reduce the entropy of the password, at least for an attacker with access to the values stored in the card. Furthermore, once the password is guessed, the scheme offers no protection against other attacks, from simply clonning the card and impersonating the user, to recovering every session key established using the password. We elaborate on this in what follows.

\subsection{Poor reparability}
One particularly weak feature of Song's scheme is that the \emph{same} key, namely $K_A = h(ID^x~\textrm{mod}~p)$, is always used to encrypt $(R_A \oplus T_A)$ during the login phase, regardless of the protocol session and during the entire life of the smart card. In general terms, this is not a recommendable practice, as it makes difficult to restore the security offered by the protocol when the user suspects that the password has been compromised.

To further clarify this, suppose that an attacker has successfully guessed the password as described above. With the information learnt she can now obtain the card's long-term secret $h(ID^x~\textrm{mod}~p)$, which could be used to fabricate a clonned card, perhaps with a different password. Even if the legitimate user suspects that the password may have been guessed, changing it does not alleviate the situation, as the same key will still be used regardless of the new password chosen! Therefore, the attacker can still impersonate the user as well as get access to future sessions keys.

The only mechanism available to the user to recover from the fact that a password has been compromised is registering again with the server \emph{using a different identity and cancelling the current one}, which is clearly unacceptable.

\subsection{Lack of perfect forward secrecy}
A trivial consequence of using the same encryption key across sessions is that the scheme does not offer perfect forward secrecy\footnote{A key establishment protocol is said to offer perfect forward secrecy if the disclosure of one secret does not compromise previously established sessions \cite{DOW, J96}}. Once $K_A$ is obtained (e.g., by guessing the password once), all previously established sessions keys can be easily computed, irrespective of the password used in past.

\subsection{Exploitation of incremental hash functions}
During the last part of the protocol, the server sends to the user the value $C_S = h(ID_A \parallel R_A' \parallel T_S)$, along with $ID_A$ and $T_S$. This construction may be extremely dangerous if $h$ is an incremental hash function (e.g., Merkle-Damg\aa rd \cite{HAC}) without a convenient finalization stage. (We note that the majority of current standarized cryptographic hash algorithms fall in this category.) If such is the case, an attacker can intercept the message and, using $C_S$ and $T_S$, go backwards through the hash algorithm and recover the internal state exactly at the point where the input $(ID_A \parallel R_A')$ has just being processed. Now, the attacker can choose a slightly different timestamp, say $T_S^{att}$, such that it will still be acceptable for the user (for example, $T_S^{att} = T_S \pm \epsilon$, with $\epsilon$ a small quantity). Using the previously recovered internal state, the attacker can compute a new value $C_S^{att} = h(ID_A \parallel R_A' \parallel T_S^{att})$, which will be forwarded to the user along with $T_S^{att}$. Note that the user cannot detect the forgery as long as the timestamp is valid, so $C_S^{att}$ will be accepted as a proof of having obtained the previously sent nonce $R_A$. However, both user and server will compute different values for the session key (as it depends on $T_S$) and they will not be able to further communicate securely even though the protocol has finished correctly.

\section{Conclusions}
We have presented an off-line password guessing attack on Song's protocol, and shown that it also has some other weaknesses despite its designer's claims. Unfortunately, being insecure seems to be the common denominator of the vast majority of the schemes proposed to date. As in the case of some other related areas, more detailed security analyses need to be performed.


\begin{thebibliography}{99}

\bibitem{DOW}
W. Diffie, P.C. van Oorschot, M.J. Wiener. ``Authentication and authenticated key exchanges.'' \emph{Designs, Codes and Cryptography} 2 (2) (1992) 107--125.

\bibitem{Haller94}
N. Haller. ``The S/KEY one-time password system.'' \emph{Proceeding of the ISOC Symposium on Network and Distributed System Security}, 1994, pp. 151--157.

\bibitem{J96}
D.P. Jablon. ``Strong Password-Only Authenticated Key Exchange.''
\emph{ACM Computer Communication Review} 26 (5) (1981) 5--26,


\bibitem{KFD}
P. Kocher, J. Jaffe, B. Jun. ``Differential power analysis.'' \emph{CRYPTO'99}, pp. 388--397.

\bibitem{Lamport81}
L. Lamport. ``Password authentication with insecure communication.'' \emph{Communications of the ACM} 24 (11) (1981) 770--772.

\bibitem{HAC}
A.J. Menezes, P.C. van Oorschot, and S.A. Vanstone. \emph{Handbook of Applied Cryptography}, Chapter 9. CRC Press, 2001.

\bibitem{MDS}
T.S. Messerges, E.A. Dabbish, R.H. Sloan. ``Examining smart-card security under the threat of power analysis attacks.'' \emph{IEEE Transactions on Computers} 51 (5) (2002) 541--552.

\bibitem{Song10}
R. Song. ``Advanced smart card based password authentication protocol.'' \emph{Computer Standards \& Interfaces} (2010), doi:10.1016/j.csi.2010.03.008.

\bibitem{XZF09}
J. Xu, W.-T Zhu, and D.-G Feng. ``An improved smart card based password authentication scheme with provable security.'' \emph{Computer Standards \& Interfaces} 31 (2009) 723--728.

\end{thebibliography}
\end{document}